\documentclass[prb,preprint,showpacs,superscriptaddress]{revtex4}
\usepackage{graphicx}

\begin{document}
\newcommand{\figwid}{12cm}
\newcommand{\cm}{cm$^{-1}$}
\newcommand{\sig}{$\sigma(\omega)$}
\newcommand{\rp}{$R_p$}
\newcommand{\bb}{K$_{0.3}$MoO$_{3}$}
\newcommand{\ab}{A$_{0.3}$MoO$_{3}$}
\newcommand{\drr}{$\Delta R/R$}
\newcommand{\tcdw}{$T_\mathrm{CDW}$}
\newcommand{\inam}{$I^{norm}_{AM}$}
\title{Coherent amplitudon generation in \bb~ through ultrafast inter-band quasi particle decay}
\author{D.~M.~Sagar}
\author{A.~A.~Tsvetkov}
\author{D.~Fausti}
\affiliation{Material Science Center, University of Groningen,
9747 AG Groningen, The Netherlands.}
\author{S. van Smaalen}
\affiliation{Laboratory of Crystallography, University of
Bayreuth, 95440 Bayreuth, Germany.}
\author{P.~H.~M.~van~Loosdrecht}
\email{P.H.M.van.Loosdrecht@rug.nl}
\affiliation{Material Science Center, University of Groningen, 9747 AG Groningen, The Netherlands.}%
\date{\today}
\begin{abstract}
The charge density wave system \bb\ has been studied  using variable
energy pump-probe spectroscopy, ellipsometry, and inelastic light
scattering. The observed transient reflectivity response exhibits
quite a complex behavior, containing contributions due to quasi
particle excitations, coherent amplitudons and phonons, and heating
effects. The generation of coherent amplitudons is discussed in
terms of relaxation of photo-excited quasi particles, and is found
to be resonant with the interband plasmon frequency. Two additional
coherent excitations observed in the transients are assigned to
zone-folding modes of the charge density wave state.
\pacs{78.47.+p,73.20.Mf,71.45.Lr,78.67.-n,78.30.-j}

\end{abstract}
\maketitle
\section{introduction}

A Charge density wave transition is a metal to insulator transition
originating from the inherent instability of a one dimensional
charge system coupled to a three dimensional
lattice\cite{Pei30,Pei55,Gru88,Gru94}. Due to the electron-phonon
coupling the electron density condenses in a charge modulated state
(modulation wavelength $\lambda=\pi/k_F$, with $k_F$ the Fermi
wavevector), and a charge density wave gap opens in the single
particle excitation spectrum at the Fermi energy. Above a certain
temperature (the Peierls temperature $T_p$) these materials are
quasi one dimensional metals, below $T_p$ they are either insulators
or semi-metals. Charge density wave systems exhibit a number of
intriguing phenomena, ranging from Luttinger liquid like behavior in
the metallic state\cite{Wan06} to highly non-linear conduction and
quasi periodic conductance oscillations in the charge ordered
state\cite{Fle78,Gru94}. The non-linear conduction seems to be a
property which is not unique to charge density wave systems, as it
is observed in other low dimensional charge ordering systems as
well\cite{Sir06}. One of the well known inorganic systems exhibiting
charge density wave transitions are the blue bronzes\cite{Sch86}.
The term bronze is applied to a variety of crystalline phases of the
transition metal oxides. They have a common formula \ab, where the
alkali metal A can be K, Rb, or Tl, and are often referred to as
{\em blue} bronzes because of their deep blue color. The crystal
structure of blue bronze contains rigid units comprised of clusters
of ten distorted MoO$_{6}$ octahedra, sharing corners along the
monoclinic b-axis \cite{Tra81}. This corner sharing provides an easy
path for the conduction electrons along the [102] directions. The
band filling is 3/4 \cite{Gru88}. The particular material addressed
in this paper is \bb\ which is a quasi-one-dimensional metal which
undergoes the metal to insulator transition (\tcdw\ = 183 K) through
the Peierls channel.\cite{Tra81,Pou83}

Apart from the usual single particle excitations (quasi particles),
charge density wave systems possess two other fundamental
excitations, which are of a collective nature. They arise from the
modulation of the charge density
$\rho(r)=\rho_0+\rho_1\cos(2k_Fr+\phi)$, and are called phasons and
amplitudons for collective phase ($\phi$) and amplitude ($\rho_1$)
oscillations, respectively. Ideally, the phason a is the gapless
Goldstone mode, leading to the notion of Fr\"ohlich
superconductivity\cite{Fro54}. However, due to the electrostatic
interactions of the charges with the underlying lattice and possibly
with impurities and imperfections, the translational symmetry of the
state is broken leading to a finite gap in the phason dispersion
spectrum. In contrast the amplitudon has an intrinsic gap in its
excitation spectrum \cite{Lee74}. In centrosymmetric media the
phason has a {\em ungerade} symmetry, and is therefore infrared
active, and the amplitudon mode (AM) has a gerade symmetry and hence
Raman active\cite{Gru88,Deg91}. The phason mode is relatively well
studied by for instance neutron scattering\cite{Esc87,Hen92} and far
infrared spectroscopy\cite{Gru88,Deg91}, and plays an important role
in the charge density wave transport. The amplitudon, {\em i.e.} the
transverse oscillation of the coupled charge-lattice system, has
been observed experimentally in for instance Raman experiments
\cite{Tra81}. Transient experiments have proven to be versatile
tools in studies on the properties of CDW materials as well.
Optically induced transient oscillatory conductivity experiments
have shown that one can increase the coherence length of the CDW
correlations by exciting quasi particles from the CDW
condensate\cite{Loo02}. An important breakthrough was the
observation that one can coherently excite the amplitudon mode in
pump-probe spectroscopy experiments\cite{Dem99,Dem02,Tsv03}. These
experiments open the possibility to study the temporal dynamics of
the collective and single particle charge density wave excitations,
as well as their interactions with quasi particles and vibrational
excitations. Demsar \textit{et al.,}\cite{Dem99} observed the
amplitudon in \bb~ as a real-time coherent modulation of the
transient reflectivity with the frequency of the amplitudon mode.
The frequency and the decay time of the AM oscillation was measured
as 1.67 THz and 10 ps, respectively. It was also found that the
single particle excitations across the CDW gap appear as a rapidly
decaying contribution to the transient reflectivity. The mechanism
of the coherent AM generation was speculated to be Displacive
Excitation of Coherent Phonons (DECP \cite{Zei92}) which is the
mechanism which describes the generation of coherent phonons in
absorbing media \cite{Val91,Val94,Gan97}. The experiment was
performed with the pump and the probe wavelength fixed at 800 nm and
at a low pump fluency of 1 $\mu$J/cm$^{2}$. The aim of the present
study is to obtain a better understanding of the transient response
of the charge density wave material \bb, and more in particular of
the generation and dephasing mechanisms of coherent amplitudon
oscillations. These issues are addressed using variable energy
pump-probe spectroscopy, ellipsometry and Raman scattering
experiments.

\maketitle
\section{Experimental}

A regenerative Ti:Sapphire amplifier seeded by a mode-locked
Ti:Sapphire laser was used to generate laser pulses at 800 nm with a
temporal width of 150 fs, operating at 1 kHz repetition rate. The
relatively low repetition rate minimizes heating effects due to the pile-up of the
pulses. To obtain laser pulses of continuously tunable energies a
Traveling Wave Optical Parametric Amplifier of Superfluorescence
(TOPAS) was used.  The TOPAS contains a series of nonlinear crystals
based on ultrashort pulse parametric frequency converters that allows
for a continuous tuning over a wide wavelength range. The output of
TOPAS is used as the pump-pulse and the wavelength of the probe is
kept at 800 nm through out the experiment. The pump and the probe
were focused to a spot of 100 microns and 50 microns diameter
respectively. In the wavelength dependent measurements the
pump-fluency was kept constant and in the pump-fluency-dependent
measurement the wavelength of the pump was kept constant. The
polarization of the incident pump pulse was parallel to the b-axis
along which the charge density wave ordering develops. The
experiments are performed in a reflection geometry with the angle of
the pump and probe pulses close to normal incidence with respect to the
sample surface. The sample was placed in a He-flow cryostat, which
allows to vary the temperature between 4.2 and 300 K (stability $\pm$0.1 K).
Raman experiments have been performed using a standard triple grating
spectrometer, using 532 nm excitation. The ellipsometry experiments have
been performed using a Woollam spectroscopic ellipsometer with the sample
placed in a special home build UHV optical cryostat.

\section{Transient reflectivity}

\begin{figure}[ht]
  \centerline{\includegraphics[width=7.5cm,clip=true]{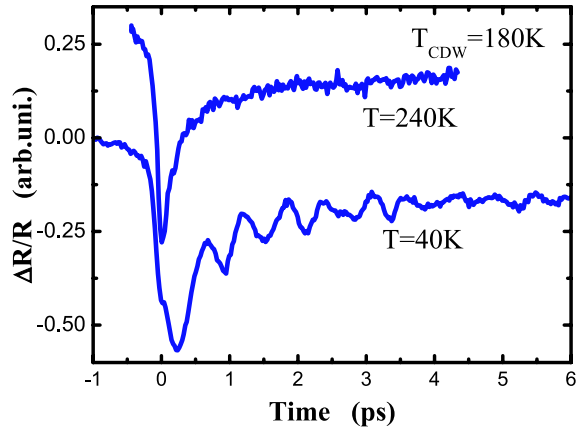}}
  \caption{\label{tcdrr}
    The transient reflectivity response of \bb\ above and below \tcdw.
  }
\end{figure}

\begin{figure}[ht]
  \centerline{\includegraphics[width=7.5cm,clip=true]{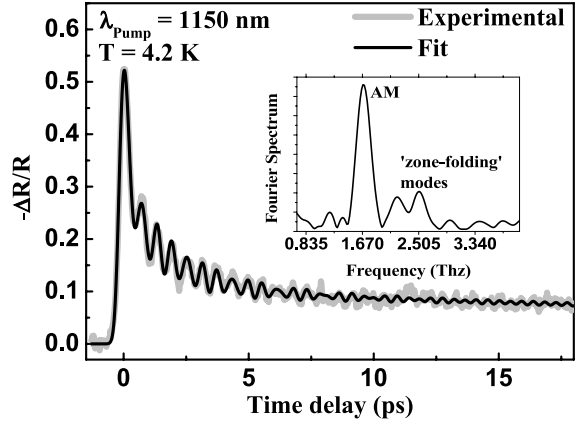}}
  \caption{\label{reflectivity}
    The transient reflectivity response of \bb\ at $T=4.2$~K, using a 1150 nm pump pulse (grey line).
    The dark line represents a fit of Eq. 1 to the data.
    {\it Inset:} Fourier spectrum showing the amplitudon mode at 1.67~THz, and two zone
    folded phonons at 2.25~THz and 2.5~THz.}
\end{figure}

\begin{figure}[ht]
  \centerline{\includegraphics[width=7.5cm,clip=true]{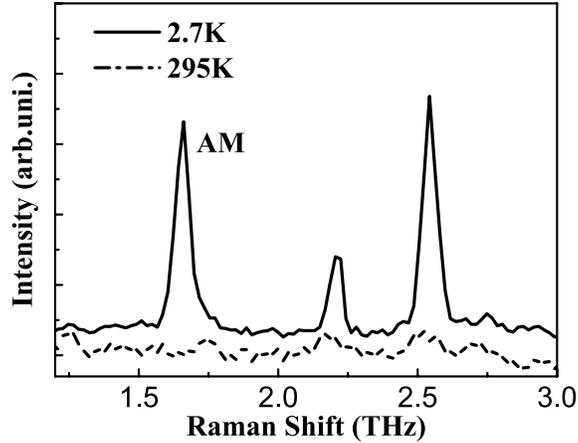}}
  \caption{\label{raman}
  The (bb) polarized raman spectrum of \bb\ above and below the CDW transition temperature,
  showing the appearance of the amplitudon mode and the two fully symmetric modes.
  }
\end{figure}

The formation of the charge density wave in blue bronze leads to
drastic changes in the nature the transient reflectivity. This is
exemplified in Fig.\ref{tcdrr} which shows two representative
transient reflectivity traces recorded above and below
\tcdw~obtained using 800 nm for both the pump and the probe
wavelengths. Above the phase transition (T = 240 K curve) the
material is metallic and gapless leading to a featureless very fast
decay (faster than the time resolution $~$150 fs) of the excited
electrons, followed by a slower decay which may be attributed to
electron-phonon coupling induced heating effects. In the charge
density wave state (Fig. \ref{tcdrr} T = 40 K curve, and Fig
\ref{reflectivity}) the response is more interesting due to the
presence of various coherent excitations and a slowing down of the
decay of the excited quasi particles resulting from the opening of
the CDW gap in the electronic excitation spectrum. A Fourier
analysis of the response (see inset Fig. \ref{reflectivity}) shows
the presence of three coherent excitations. The strongest component
found at 1.67 THz has been attributed to the coherent excitation of
the collective amplitudon mode \cite{Dem99}. The two additional
modes at 2.25 and 2.5 THz can be attributed to Raman active phonons
which are activated in the CDW state due to folding of the Brillouin
zone \cite{Pou83}. Polarized frequency domain Raman scattering
experiments have indeed confirmed this interpretation (see Fig.
\ref{raman}). The excitation wavelength for this experiment was 532
nm with the polarization of the incoming and scattered light
parallel to the b-axis of the crystal. Above the phase transition
temperature (T = 295 K spectrum) the Raman spectrum is rather featureless
in the region of the amplitudon mode. In contrast, the 2.7 K
spectrum shows the appearance of just the three modes which are also
observed in the time domain traces.

The electronic contribution to \drr~can be discussed in terms of
their relaxation time scales. Since the energy of the pump pulse is
much larger than the CDW gap $(\Delta_{CDW}$=0.12 eV \cite{Deg91})
one expects that the optical pumping excites a large number of quasi
particles ($\frac{E_{Pump}}{\Delta}$~$\sim$ 30-50 QP's per photon)
across the CDW-gap. After this photo excitation a very fast internal
thermalization of the highly non-equilibrium electron distribution
occurs on a time scale of a few fs, which is followed by
electron-phonon thermalization that occurs on a time scale less than
100 fs \cite{Dem99}. Although the temporal resolution of the current
experiments is not enough to observe these effects, once the quasi
particles have internally thermalized and relaxed to states near the
Fermi level, further energy relaxation is delayed due to the
presence of the CDW-gap. The presence of the CDW-gap acts as a
"bottle-neck" for the thermalized quasi particles
\cite{bil93,odi01}. The typical time for relaxation over the CDW gap
is found to be 0.6 ps in the present experiments. However, the
observed decay is not a simple exponential but it is rather a
stretched exponential decay which is typical for a system with a
distribution of relaxation times as is for instance found in systems
with an anisotropic gap such as 1{\it T}-TaS$_2$\cite{Dem02}. Even
though blue bronze is only quasi-one dimensional, there is no
evidence for an anisotropic gap in this system. It is more likely
that the stretched decay originates from distribution of relaxation
times resulting from a possible glassy nature of the CDW ground
state\cite{bil93,odi01}. Finally, like in the high temperature
phase, a long time relaxation of \drr\ is observed, which may be
attributed to heating effects. No qualitative difference with the
high temperature relaxation is observed here, ruling out a possible
contribution of phasons to the response as has been suggested
earlier \cite{Dem99}.

To summarize, the observed transient reflectivity response in the
CDW phase can be described by
\begin{equation}\label{eq:one}
\frac{\Delta R}{R}= A e^{(-t/\tau_{QP})^n} + \sum_j A_j~
e^{-t/\tau_j} \cos(\omega_j t) + B e^{-t/\tau_{L}}\ \ \ ,
\end{equation}
where the first term describes the quasi particle response using a
stretched exponential decay with time constant $\tau_{QP}$ and
stretch index $n$ which takes the relaxation of the quasi particles
across the CDW gap into account. The second term in this expression
accounts for the observed coherent amplitudon and phonon
oscillations with frequencies $\omega_j$ and decay times $\tau_j$,
and the last term represents the observed long time response with a
decay time $\tau_L$ presumably originating from heating effects.

A fit of Eq.~\ref{eq:one} to the data generally leads to excellent
agreement with the data, as is for instance shown in
Fig.~\ref{reflectivity}. For this fit the quasi particle decay time is found
to be $\tau_{QP}=0.65$~ps with $n=1/2$, and the amplitudon lifetime
$\tau_{AM}=3.5$~ps. The decay times of the two coherent phonons is
of the order of 20$\pm$5 ps, whereas the cooling time is to slow to
determine with any accuracy ($\tau_{L}>60$~ps). It is interesting to
note that the coherent amplitudon is very short lived when compared
to the two coherent phonons. This heavy damping of coherent
amplitudon presumably results from amplitudon-quasi particle
scattering leading either to a decay or to a dephasing of the
coherent amplitudon excitation. The observation that the decay time
in the present experiment is shorter than in the experiments by
Demsar et al,\cite{Dem99}, who reported $\tau_{AM}=10$~ps is
consistent with this. The density of excited quasi particles in the
current experiments is substantially higher (several orders of
magnitude) than in the previous experiments,
leading to this faster decay of the coherent amplitudon mode,
and also to a larger magnitude of the induced response
(here as large as 0.1, compared to 10$^{-4}$ in
Ref.\cite{Dem99}). The quasi particle lifetime measured for the
lowest pump fluency is about 0.6 ps which is close to the value
measured by Demsar et al. (0.5 ps) \cite{Dem99}. Besides the
quasi particle density, the main difference is that the experimental
temperature in our case is T = 4.2 K whereas in \cite{Dem99} it is T
= 45 K. The stretched exponential behavior observed in the present
study, in contrast to the single exponent in Demsar's study, might
be due to the presence of a glassy state which has recently been
discussed in literature\cite{bil93,odi01,Star04}.
The stretched exponential behavior is typical for a glassy state.

\section{Pump fluency dependence\label{sec:fluency}}

\begin{figure}[hbt]
  \centerline{\includegraphics[width=7.5cm,clip=true]{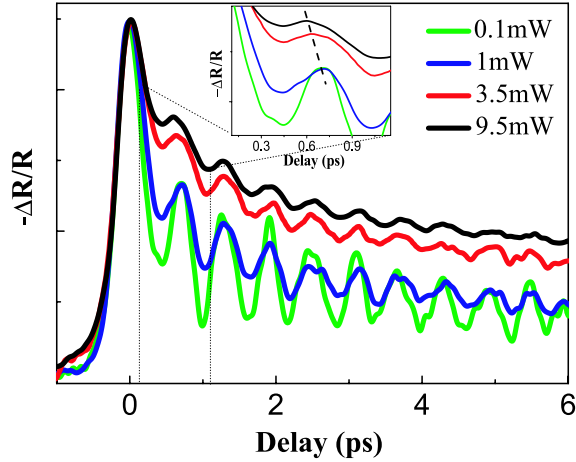}}
  \caption{\label{pumpdep}(Color online)
Normalized time resolved reflectivity traces for 800 nm pump-probe experiments
with a pump fluency varying between 0.1 and 10 mW. The data is normalized to
the zero delay response. The insert shows
the position of the first beating of the coherent mode evidencing a phase shift
upon increasing fluency.}
\end{figure}

\begin{figure}[hb!]
  \centerline{\includegraphics[width=7.5cm,clip=true]{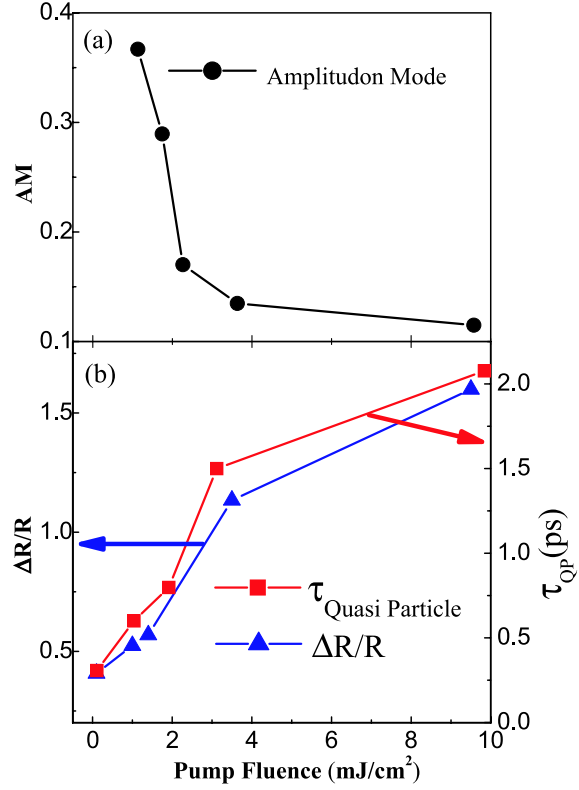}}
  \caption{\label{parameter}(Color online)
  Dependence of the AM amplitude (a), and the quasi particle lifetime and amplitude of the quasi particle
  peak (b) on the pump fluency. The QP life time and the quasi particle peak show a linear increase with pump fluency, while the AM amplitude shows a dramatic decrease. The solid lines are guides to the eye.}
\end{figure}

In order to address the generation mechanism of the coherent AM
pump-fluency dependent (this section) and pump-wavelength (see section \ref{wave})
dependent experiments have been carried out.
The experiments reported in Fig. \ref{pumpdep} are performed with a
relatively high pump fluency of 1-10 mJ/cm$^2$. Note that the data
are scaled to the strength of the initial response, {\em i.e.} to the
strength of the quasi particle peak. The strength of the quasi particle peak itself
is found to be linearly dependent on the pump fluency (see Fig. \ref{parameter} (b)).
This linear dependence is consistent with
the expectation that the density of the excited electrons scales
with the mean number of photon of the light pulse. Hence, if the
generation mechanism of the coherent AM modes can be described, as
suggested in \cite{Demsar1}, by the simple theory of displacive
excitation (DECP) \cite{Zei92}, a linear dependence of the coherent
excitation amplitude with the pump fluency is expected. On the
contrary, Fig.\ref{pumpdep} and Fig.\ref{parameter}(a) clearly show that the intensity of
the coherent amplitudon oscillations decreases dramatically with
increasing the pump fluency. This immediately rules out the
possibility of a linear relationship between the number of photons
of the pump pulse and the observed amplitude of the coherent AM mode.
A second striking feature of the pump fluency dependence of the
time resolved reflectivity traces is the substantial increase in the
quasi particle lifetime with the pump power (see Fig.\ref{parameter}(b)).

The increasing lifetime of the quasi particles, which is probably
due to state filling effects, together with the decreasing
amplitudon amplitude seems to indicate that the coherent amplitudons
are not directly generated by a coupling to the photons, but rather
by a coupling to the decay of the quasi particle excitations. {\em
I.e.} the coherent amplitudon oscillation are generated by the decay
processes of the excited electron population. The increasing of the
quasi particle lifetime leads to a reduction of the coherence of the
amplitudons generated, and thereby to a decrease in the amplitude of
the coherent response. In a very simple approach, neglecting the
temporal width of the pump-pulse and the stretched exponential
behavior of the decay of the quasi particle, one can model the quasi
particle response by a simple exponential decay
($e^{-t/\tau_{QP}}/\tau_{QP}$ for $t>0$), and a linear coupling
between the quasi particle decay and the amplitudon generation. The
resulting coherent amplitudon response is then given by:
\begin{equation}\label{AM-Size}
\int_{0}^{\infty}\frac{A}{\tau_{QP}}~e^{-t'/\tau_{QP}}cos(\Omega(t+t'))=\frac{A}{1+\Omega_{AM}^{2}\tau_{QP}^{2}}cos(\Omega
t+arctan(\Omega\tau_{QP}))
\end{equation}
where $A$ is the integrated area of the quasi particle peak, $\Omega$
is the frequency of the AM mode, and $\tau_{QP}$ is the quasi particle
lifetime. The product $\Omega_{AM}\tau_{QP}$ plays the role of a decoherence factor.
For this special case, the dependence of the
size of the coherent amplitudon response on $\Omega_{AM}\tau_{QP}$ is
Lorentzian. This is valid as long as the quasi particle response is slower than the temporal pump pulse width.
For short quasi particle relaxation times, the quasi particle response becomes more symmetric, leading to
an decaying exponential dependence of the coherent signal on $\Omega_{AM}\tau_{QP}$.

\begin{figure}[hb!]
  \centerline{\includegraphics[width=7.5cm,clip=true]{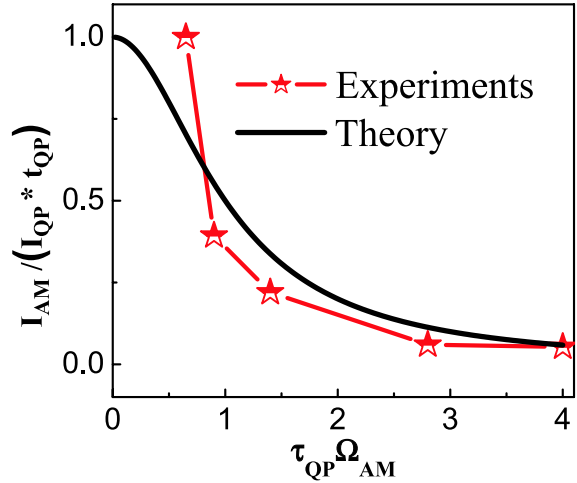}}
  \caption{\label{decaytime}(Color online)
  Scaled amplitude (see text) of the coherent AM as a function of the decoherence factor $\tau_{QP}\Omega_{AM}$ (symbols).
  The decrease with increasing quasi particle life time shows the
  loss of coherence due to the delayed relaxation of the quasi particles. The solid line
  displays the amplitude of Eq. \ref{AM-Size} as a function of the coherence factor.}
\end{figure}

To elucidate the point, Fig.\ref{decaytime} shows the amplitude of the
coherent amplitudon response, normalized to the quasi particle response,
as a function of the decoherence factor (symbols).
The quasi particle response is approximated as the product of
its intensity and lifetime for a given fluency.
Inspection of Fig.\ref{decaytime} shows that
the intensity of the coherent amplitudon decreases rapidly as the
decoherence factor reaches unity, beyond which the coherence is lost
due to the increase of the quasi particle lifetime and hence dephasing
of the generated amplitudons. The same figure also shows the amplitude
obtained from Eq.\ref{AM-Size}.
Given the simplicity of the model and the fact that
there are no adjustable parameters, the agreement
between the experimental data and Eq.\ref{AM-Size} is rather striking.
The small deviation for low decoherence factor is probably due to the assumption
of an exponentially decaying quasi particle population. This can
be improved by using a more realistic quasi particle response.

In addition to this the decreasing size of the coherent response,
the model proposed by (Eq. \ref{AM-Size}) also predicts
a phase shift of the coherent amplitudon response proportional to
$\arctan(\Omega_{AM}\tau_{QP})$.
This is indeed what is experimentally observed.
Comparing just the lowest and the highest fluency (with $\tau_{QP}=$~0.3 and 2.0 ps, respectively),
the expected phase shift is $\simeq2\pi/8$, which is consistent with the experimentally
observed shift (see insert Fig. \ref{pumpdep}).

Some further evidence for the coupling of the photo excited
quasi particles to the collective amplitudon mode of the charge
density wave state may be found from the pump-wavelength dependent
experiments discussed in the next section.

\section{The Wavelength Dependence\label{wave}}

Before discussing the pump-wavelength dependent transient CDW
dynamics it is instructive to consider the linear optical
response of the system. The linear optical response at various
wavelengths provides information on possible absorption bands of
the material, eventually giving more insight into the coupling of
the electronic transitions with the CDW excitations.

\begin{figure}[ht]
  \centerline{\includegraphics[width=7.5cm,clip=true]{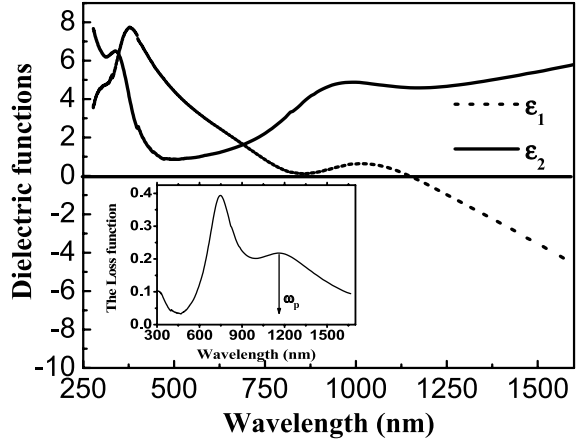}}
  \caption{\label{dielectric}
   Optical response $\epsilon_{1}$ and $\epsilon_{2}$ of \bb. {\it Inset:}
   The energy loss function calculated from the data in the main panel.}
   \end{figure}

Fig.\ref{dielectric} shows the optical response in the
wavelength region ranging from 270 nm to 1700 nm.
The sharp band in $\epsilon_{2}$ raising from 500 nm toward the
lower wavelength side is due to the "p-d"-transitions involving the
electronic excitations from the Oxygen "2p" to the Molybdenum "4d"
levels. The quotations are used to indicate that the levels are
admixtures rather than pure ones. This is consistent with the
photo-emission and electron energy loss experiments done on
blue bronze \cite{Wer85,Sin99}. The broad asymmetric band
around 1000 nm is due to interband "d-d"-transitions.
The actual zero crossing of $\epsilon_{1}$ occurs at 1150 nm (1.08 eV).
This, however, does not correspond to the actual plasma energy,
which occurs at 1.5 eV\cite{Sin99}. This is also demonstrated in
the inset of Fig.\ref{dielectric}, which shows the energy loss spectrum
calculated from the dielectric function. This spectrum shows two
peaks, one at the actual plasma frequency (where also a distinct minimum
in $\epsilon_1$ is observed), and one at 1.08 eV.
This latter energy corresponds to an interband plasmon involving
the Mo-d derived dispersionless conduction branch predicted by
tight-binding calculations of Travaglini and Wachter\cite{Tra85,Sin99}.

\begin{figure}[ht]
  \centerline{\includegraphics[width=7.5cm,clip=true]{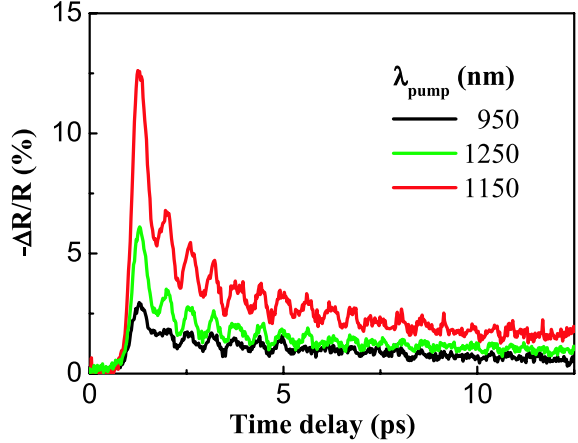}}
  \caption{\label{dRbyw}(Color online)
  Variation of the transient response with the pump wavelength.}
\end{figure}
The transient response, and in particular the coherent amplitudon
response, depends strongly on the excitation wavelength, and is
found to be strongest for $\lambda_{Pump}\sim 1150$~nm and at small
wavelengths. This is demonstrated in Fig.\ref{dRbyw}, which shows
the transient response for few selected pump wavelengths
($\lambda_{Pump}$).

\begin{figure}[ht]
  \centerline{\includegraphics[width=7.5cm,clip=true]{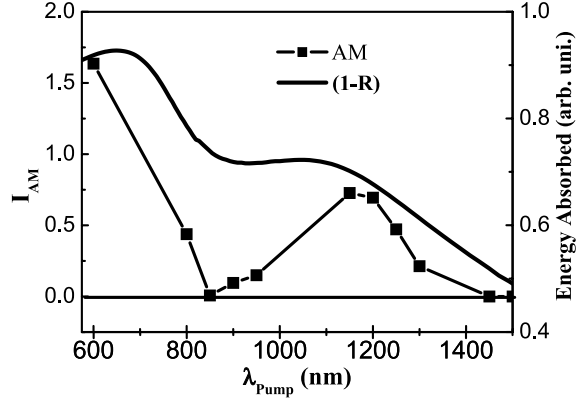}}
  \caption{\label{amplitude}
  Coherent amplitudon response as a function of the pump wavelength (symbols, the
  drawn line is a guide to the eye).
  The total absorbed pump energy 1-R, with R the normal incidence reflectivity,
  is also plotted for comparison (solid line).
  }
\end{figure}

Overall, the amplitude of the coherent amplitudon mode follows the
pump energy absorbed in the material. This can be seen from
Fig.\ref{amplitude}, which shows the amplitude of the coherent
amplitudon mode measured for various pump wavelengths at a constant
pump fluency of 1 mJ/cm$^{2}$. The same graph also shows the amount
of absorbed pump energy calculated from the optical data in
Fig.\ref{dielectric}. It is found that the quasi particle lifetime
does not vary strongly with the wavelength. This, together with the
fact that the efficiency of the coherent amplitudon generation
roughly follows the amount of absorbed energy, and hence the number
of photo-excited quasi particles, suggests once again the quasi
particle induced nature of the coherent excitation.

Although the coherent AM roughly follows the absorbed energy curve,
it does actually not exactly scale with it. In particular, the AM
response (as well as the quasi particle response) is markedly peaked
near 1150 nm, corresponding to the interband plasma wavelength
discussed above. This is surprising since one does not expect the
light to couple directly to the plasmon modes (as can also be seen
from Fig.\ref{dielectric}). Nevertheless, the experiments do
evidence an efficient coupling most likely through highly excited
quasi particles which relax by the emission of plasmon excitations,
or though a coupling to surface plasmons. Once excited, the
interband plasmons can relax either by emission of lower energy
quasi particles which subsequently can excite amplitudons, as the
enhanced quasi particle response seems to suggest, or possibly even
directly via decay into amplitudon modes.

\section{Conclusion}
In summary, we studied \bb\ using time resolved spectroscopy,
ellipsometry, and polarized Raman spectroscopy. The transient
reflectivity experiments show, in addition to the coherent
amplitudon mode observed previously\cite{Dem99}, two coherent phonon
modes which are also observed in the low temperature Raman spectra.
They are assigned to zone-folding modes associated with the charge
density wave transition. The lifetime of the coherent amplitudon
mode is found to be relatively short, which is believed to be due to
the coupling of the AM to the high density of quasi particles.

The generation mechanism of coherent amplitudons in blue bronze does not seem
to be the usual DECP\cite{Zei92} mechanism, nor the transient stimulated
Raman mechanism\cite{Mer96}, since these are not consistent with the
observed non linear pump power dependence of the magnitude of the response.
It is found that the magnitude of the coherent amplitudon mode
depends on the dynamics of the quasi particle decay.
It is therefore beleived that the coherent amplitudons are generated
through the decay of quasi particles over the Peierls gap. A simple model
taking based on this notion accounts well for the observed non linear
pump power dependence, as well as for the observed phase shift in
the coherent amplitudon response.  Pump wavelength dependent experiments,
where the coherent amplitudon response is found to be consistent with
the equilibrium absorption derived by ellipsometry measurements (and hence
with the quasi particle response) further supports the proposed interpretation.
The mechanism proposed could be relevant in the generation mechanism
of coherent excitations in other highly absorbing materials as well. Further
investigations of different compounds are necessary to confirm this.



\end{document}